\catcode`\@=11

\font\tenmsa=msam10
\font\sevenmsa=msam7
\font\fivemsa=msam5
\font\tenmsb=msbm10
\font\sevenmsb=msbm7
\font\fivemsb=msbm5
\newfam\msafam
\newfam\msbfam
\textfont\msafam=\tenmsa  \scriptfont\msafam=\sevenmsa
  \scriptscriptfont\msafam=\fivemsa
\textfont\msbfam=\tenmsb  \scriptfont\msbfam=\sevenmsb
  \scriptscriptfont\msbfam=\fivemsb

\def\hexnumber@#1{\ifnum#1<10 \number#1\else
 \ifnum#1=10 A\else\ifnum#1=11 B\else\ifnum#1=12 C\else
 \ifnum#1=13 D\else\ifnum#1=14 E\else\ifnum#1=15 F\fi\fi\fi\fi\fi\fi\fi}

\def\msa@{\hexnumber@\msafam}
\def\msb@{\hexnumber@\msbfam}
\mathchardef\boxdot="2\msa@00
\mathchardef\boxplus="2\msa@01
\mathchardef\boxtimes="2\msa@02
\mathchardef\square="0\msa@03
\mathchardef\blacksquare="0\msa@04
\mathchardef\centerdot="2\msa@05
\mathchardef\lozenge="0\msa@06
\mathchardef\blacklozenge="0\msa@07
\mathchardef\circlearrowright="3\msa@08
\mathchardef\circlearrowleft="3\msa@09
\mathchardef\rightleftharpoons="3\msa@0A
\mathchardef\leftrightharpoons="3\msa@0B
\mathchardef\boxminus="2\msa@0C
\mathchardef\Vdash="3\msa@0D
\mathchardef\Vvdash="3\msa@0E
\mathchardef\vDash="3\msa@0F
\mathchardef\twoheadrightarrow="3\msa@10
\mathchardef\twoheadleftarrow="3\msa@11
\mathchardef\leftleftarrows="3\msa@12
\mathchardef\rightrightarrows="3\msa@13
\mathchardef\upuparrows="3\msa@14
\mathchardef\downdownarrows="3\msa@15
\mathchardef\upharpoonright="3\msa@16

\mathchardef\downharpoonright="3\msa@17
\mathchardef\upharpoonleft="3\msa@18
\mathchardef\downharpoonleft="3\msa@19
\mathchardef\rightarrowtail="3\msa@1A
\mathchardef\leftarrowtail="3\msa@1B
\mathchardef\leftrightarrows="3\msa@1C
\mathchardef\rightleftarrows="3\msa@1D
\mathchardef\Lsh="3\msa@1E
\mathchardef\Rsh="3\msa@1F
\mathchardef\rightsquigarrow="3\msa@20
\mathchardef\leftrightsquigarrow="3\msa@21
\mathchardef\looparrowleft="3\msa@22
\mathchardef\looparrowright="3\msa@23
\mathchardef\circeq="3\msa@24
\mathchardef\succsim="3\msa@25
\mathchardef\gtrsim="3\msa@26
\mathchardef\gtrapprox="3\msa@27
\mathchardef\multimap="3\msa@28
\mathchardef\therefore="3\msa@29
\mathchardef\because="3\msa@2A
\mathchardef\doteqdot="3\msa@2B

\mathchardef\triangleq="3\msa@2C
\mathchardef\precsim="3\msa@2D
\mathchardef\lesssim="3\msa@2E
\mathchardef\lessapprox="3\msa@2F
\mathchardef\eqslantless="3\msa@30
\mathchardef\eqslantgtr="3\msa@31
\mathchardef\curlyeqprec="3\msa@32
\mathchardef\curlyeqsucc="3\msa@33
\mathchardef\preccurlyeq="3\msa@34
\mathchardef\leqq="3\msa@35
\mathchardef\leqslant="3\msa@36
\mathchardef\lessgtr="3\msa@37
\mathchardef\backprime="0\msa@38
\mathchardef\risingdotseq="3\msa@3A
\mathchardef\fallingdotseq="3\msa@3B
\mathchardef\succcurlyeq="3\msa@3C
\mathchardef\geqq="3\msa@3D
\mathchardef\geqslant="3\msa@3E
\mathchardef\gtrless="3\msa@3F
\mathchardef\sqsubset="3\msa@40
\mathchardef\sqsupset="3\msa@41
\mathchardef\trianglerighteq="3\msa@44
\mathchardef\trianglelefteq="3\msa@45
\mathchardef\bigstar="0\msa@46
\mathchardef\between="3\msa@47
\mathchardef\blacktriangledown="0\msa@48
\mathchardef\blacktriangleright="3\msa@49
\mathchardef\blacktriangleleft="3\msa@4A
\mathchardef\blacktriangle="0\msa@4E
\mathchardef\triangledown="0\msa@4F
\mathchardef\eqcirc="3\msa@50
\mathchardef\lesseqgtr="3\msa@51
\mathchardef\gtreqless="3\msa@52
\mathchardef\lesseqqgtr="3\msa@53
\mathchardef\gtreqqless="3\msa@54
\mathchardef\Rrightarrow="3\msa@56
\mathchardef\Lleftarrow="3\msa@57
\mathchardef\veebar="2\msa@59
\mathchardef\barwedge="2\msa@5A
\mathchardef\doublebarwedge="2\msa@5B
\mathchardef\angle="0\msa@5C
\mathchardef\measuredangle="0\msa@5D
\mathchardef\sphericalangle="0\msa@5E
\mathchardef\varpropto="3\msa@5F
\mathchardef\smallsmile="3\msa@60
\mathchardef\smallfrown="3\msa@61
\mathchardef\Subset="3\msa@62
\mathchardef\Supset="3\msa@63
\mathchardef\Cup="2\msa@64

\mathchardef\Cap="2\msa@65

\mathchardef\curlywedge="2\msa@66
\mathchardef\curlyvee="2\msa@67
\mathchardef\leftthreetimes="2\msa@68
\mathchardef\rightthreetimes="2\msa@69
\mathchardef\subseteqq="3\msa@6A
\mathchardef\supseteqq="3\msa@6B
\mathchardef\bumpeq="3\msa@6C
\mathchardef\Bumpeq="3\msa@6D
\mathchardef\lll="3\msa@6E

\mathchardef\ggg="3\msa@6F

\mathchardef\circledS="0\msa@73
\mathchardef\pitchfork="3\msa@74
\mathchardef\dotplus="2\msa@75
\mathchardef\backsim="3\msa@76
\mathchardef\backsimeq="3\msa@77
\mathchardef\complement="0\msa@7B
\mathchardef\intercal="2\msa@7C
\mathchardef\circledcirc="2\msa@7D
\mathchardef\circledast="2\msa@7E
\mathchardef\circleddash="2\msa@7F
\def\ulcorner{\delimiter"4\msa@70\msa@70 }
\def\urcorner{\delimiter"5\msa@71\msa@71 }
\def\llcorner{\delimiter"4\msa@78\msa@78 }
\def\lrcorner{\delimiter"5\msa@79\msa@79 }
\def\yen{\mathhexbox\msa@55 }
\def\checkmark{\mathhexbox\msa@58 }
\def\circledR{\mathhexbox\msa@72 }
\def\maltese{\mathhexbox\msa@7A }
\mathchardef\lvertneqq="3\msb@00
\mathchardef\gvertneqq="3\msb@01
\mathchardef\nleq="3\msb@02
\mathchardef\ngeq="3\msb@03
\mathchardef\nless="3\msb@04
\mathchardef\ngtr="3\msb@05
\mathchardef\nprec="3\msb@06
\mathchardef\nsucc="3\msb@07
\mathchardef\lneqq="3\msb@08
\mathchardef\gneqq="3\msb@09
\mathchardef\nleqslant="3\msb@0A
\mathchardef\ngeqslant="3\msb@0B
\mathchardef\lneq="3\msb@0C
\mathchardef\gneq="3\msb@0D
\mathchardef\npreceq="3\msb@0E
\mathchardef\nsucceq="3\msb@0F
\mathchardef\precnsim="3\msb@10
\mathchardef\succnsim="3\msb@11
\mathchardef\lnsim="3\msb@12
\mathchardef\gnsim="3\msb@13
\mathchardef\nleqq="3\msb@14
\mathchardef\ngeqq="3\msb@15
\mathchardef\precneqq="3\msb@16
\mathchardef\succneqq="3\msb@17
\mathchardef\precnapprox="3\msb@18
\mathchardef\succnapprox="3\msb@19
\mathchardef\lnapprox="3\msb@1A
\mathchardef\gnapprox="3\msb@1B
\mathchardef\nsim="3\msb@1C
\mathchardef\napprox="3\msb@1D
\mathchardef\nsubseteqq="3\msb@22
\mathchardef\nsupseteqq="3\msb@23
\mathchardef\subsetneqq="3\msb@24
\mathchardef\supsetneqq="3\msb@25
\mathchardef\subsetneq="3\msb@28
\mathchardef\supsetneq="3\msb@29
\mathchardef\nsubseteq="3\msb@2A
\mathchardef\nsupseteq="3\msb@2B
\mathchardef\nparallel="3\msb@2C
\mathchardef\nmid="3\msb@2D
\mathchardef\nshortmid="3\msb@2E
\mathchardef\nshortparallel="3\msb@2F
\mathchardef\nvdash="3\msb@30
\mathchardef\nVdash="3\msb@31
\mathchardef\nvDash="3\msb@32
\mathchardef\nVDash="3\msb@33
\mathchardef\ntrianglerighteq="3\msb@34
\mathchardef\ntrianglelefteq="3\msb@35
\mathchardef\ntriangleleft="3\msb@36
\mathchardef\ntriangleright="3\msb@37
\mathchardef\nleftarrow="3\msb@38
\mathchardef\nrightarrow="3\msb@39
\mathchardef\nLeftarrow="3\msb@3A
\mathchardef\nRightarrow="3\msb@3B
\mathchardef\nLeftrightarrow="3\msb@3C
\mathchardef\nleftrightarrow="3\msb@3D
\mathchardef\divideontimes="2\msb@3E
\mathchardef\varnothing="0\msb@3F
\mathchardef\nexists="0\msb@40
\mathchardef\mho="0\msb@66
\mathchardef\thorn="0\msb@67
\mathchardef\beth="0\msb@69
\mathchardef\gimel="0\msb@6A
\mathchardef\daleth="0\msb@6B
\mathchardef\lessdot="3\msb@6C
\mathchardef\gtrdot="3\msb@6D
\mathchardef\ltimes="2\msb@6E
\mathchardef\rtimes="2\msb@6F
\mathchardef\shortmid="3\msb@70
\mathchardef\shortparallel="3\msb@71
\mathchardef\smallsetminus="2\msb@72
\mathchardef\thicksim="3\msb@73
\mathchardef\thickapprox="3\msb@74
\mathchardef\approxeq="3\msb@75
\mathchardef\succapprox="3\msb@76
\mathchardef\precapprox="3\msb@77
\mathchardef\curvearrowleft="3\msb@78
\mathchardef\curvearrowright="3\msb@79
\mathchardef\digamma="0\msb@7A
\mathchardef\varkappa="0\msb@7B
\mathchardef\hslash="0\msb@7D
\mathchardef\hbar="0\msb@7E
\mathchardef\backepsilon="3\msb@7F
\def\Bbb{\ifmmode\let\next\Bbb@\else
 \def\next{\errmessage{Use \string\Bbb\space only in math mode}}\fi\next}
\def\Bbb@#1{{\Bbb@@{#1}}}
\def\Bbb@@#1{\fam\msbfam#1}

\catcode`\@=\active

\def\inv{^{\raise.15ex\hbox{${
  \scriptscriptstyle -}$}\kern-.05em 1}}

\def\Dsl{\,\raise.15ex\hbox{$/$}\mkern-13.5mu D}
\def\dsl{\raise.15ex\hbox{$/$}\kern-.57em\hbox{$\partial$}}

\def\lspace{\ifx\answ\bigans{}\else\qquad\fi}
\def\del{\partial}

\def\CL{\hbox{{$\cal L$}}}

 \def\CZ{\hbox{{$\cal Z$}}}

\def\grav{{\scriptscriptstyle G}}

\def\cm{\hbox{{\sl m}}} 
\def\cg{\hbox{{\sl g}}} 

\def\lform{\hbox{$\sqcup$}\llap{\hbox{$\sqcap$}}}
\def\darr#1{\raise1.5ex\hbox{$\leftrightarrow$}
\mkern-16.5mu #1}

\def\INT{{\textstyle \int\kern-.642em\int}}

\def\R{{\Bbb R}}
\def\C{{\Bbb C}}
\def\Z{{\Bbb Z}}

\def\eps{{\epsilon}}

\def\rcross{{\triangleright\!\!\!<}}
\def\lcross{{>\!\!\!\triangleleft}}
\def\rcocross{{\blacktriangleright\!\!<}}

\def\cobicross{{\triangleright\!\!\!\blacktriangleleft}}
\def\bicross{{\blacktriangleright\!\!\!\triangleleft}}
\def\dcross{{\bowtie}}

\def\rbiprod{{\cdot\kern-.33em\triangleright\!\!\!<}}
\def\lbiprod{{>\!\!\!\triangleleft\kern-.33em\cdot}}

\def\tens{\mathop{\otimes}}

\def\la{{\triangleright}}\def\ra{{\triangleleft}}

\def\extd{{{\rm d}}}

\def\swap{{\leftrightarrow}}

\def\isom{{\cong}}

\def\span{{\rm span}}

\def\ad{{\rm ad}}

\def\coev{{\rm coev}}

\def\id{{\rm id}}

\def\grav{{\scriptstyle G}}
\def\<{\langle}
\def\>{\rangle}

\def\vecs{{\bf s}}
\def\vecu{{\bf u}}\def\vecx{{\bf x}}\def\vecp{{\bf p}}
\def\veca{{\bf a}}

\def\<{\langle}
\def\>{\rangle}

\def\equad{\kern -1.7em}

\def\nquad{{\!\!\!\!\!\!}}

\def\und#1{{\underline {#1}}}

\def\text#1{\mbox{\rm #1}}
\def\note#1{}

\def\blacksquare{{\lform}}
\def\frac#1#2{{{#1\over#2}}}

\def\eqn#1#2{\begin{equation}#2\label{#1}\end{equation}}

\def\align#1{\begin{eqnarray*}#1\end{eqnarray*}}

\def\cmath#1{\[\begin{array}{c} #1 \end{array}\]}
\def\ceqn#1#2{\begin{equation}\label{#1}\begin{array}{c}#2\end{array}
\end{equation}}

\documentstyle[11pt,epsf]{article}
\newtheorem{lemma}{Lemma}[section] \newtheorem{propos}[lemma]{Proposition}

\begin{document}\baselineskip 20pt

{\ }\qquad DAMTP/94-69 to app. Proc. 1st Gursey Memorial Conf.
Strings and Symmetries, Istanbul, June 1994.
\vspace{.2in}

\begin{center} {\LARGE DUALITY PRINCIPLE AND BRAIDED GEOMERTY}
\\ \baselineskip 13pt{\ }
{\ }\\ S. Majid\footnote{Royal Society University Research Fellow and Fellow of
Pembroke College, Cambridge}\\
{\ }\\
Department of Applied Mathematics \& Theoretical Physics\\
University of Cambridge, Cambridge CB3 9EW
\end{center}

\vspace{10pt}
\begin{quote}\baselineskip 13pt
\noindent{\bf Abstract} We give an overview of a new kind symmetry in physics
which
exists between observables and states and which is made possible by the
language
of Hopf algebras and quantum geometry.  It has been proposed by the author as a
feature of Planck scale physics.  More recent work includes corresponding
results at the semiclassical level of Poisson-Lie groups and at the level of
braided groups and braided geometry.

\end{quote}
\baselineskip 20pt

\def\xpos{{x}}

\section{Introduction}

I did not know Feza G\"ursey personally, but I would guess that he understood
very well one of the principles close to my own heart:  the deep
interdependence
of theoretical physics and pure mathematics.  A fairly conventional view,
expressed neatly by Pierre Ramond in his talk, is that we theoretical
physicists
are probing into higher and higher energy scales building finer and more
accurate theories in the process.  Some say that we can learn in this task from
pure mathematicians to supply us with the right conceptual structures, others
say that they can learn from us to find examples.  My own view, however, is
more
radical but I will try nevertheless to outline it here and to reflect it in my
talk.  The starting point is that after all Nature does not care what
mathematics has been discovered until now.  So it would be rather naive to
think
that the maths that we know now is going to be enough to reach higher and
higher
levels until we reach the Planck scale or beyond to the ultimate theory of
physics.  Most likely if we did meet someone who knew this ultimate theory of
physics and asked them to tell us, we would not understand a word of it because
of {\em completely new} mathematical concepts that are centuries from being
invented.  Put another way, we physicists should not only try to apply known or
fashionable mathematics, we should be ready and willing to invent whatever we
think Nature uses as we go along in our search:  we should be equally good {\em
pure} mathematicians (in the creative sense, not necessarily in terms of
rigour)
as we are calculators and applied ones.  Sometimes one can forget that
mathematics is about creating natural ideas and concepts, and not about
theorems
and lemmas {\em per se}, though these are how a mathematician knows that the
concepts are worthwhile.  Physicists have perhaps other intuitive ways of
knowing what is worthwhile.

Put another way, the naive view is that Nature exists and is out there, and
 maths exists and is out there, and we just have to apply the latter to
describe
 the former.  By contrast, I would like to argue that the word `exists' is
 questionable in both cases, and that in fact each creates and justifies the
 other as we climb to higher and higher energy scales in our understanding.  I
 mean that Nature and Language (which for me means mathematics) define each
 other and are at the very least interdependent.

I would like to describe here how this philosophy works in my own approach to
unifying physics, which includes, as for many physicists, the goal of unifying
quantum mechanics and gravity.  It concerns a new kind of geometry more general
than Riemannian and powerful enough to not break down in the quantum domain.
Planck scale physics is the reason that I first became interested in Hopf
algebras (quantum groups) and quantum or non-commutative geometry.  It has
nothing to do with the reason that quantum groups subsequently became very
fashionable and important, which has more to do with work on inverse
scattering and applications to knot and three-manifold invariants.  It is
gratifying that the same kind of new mathematics is emerging from all these
different directions.  The goal in this lecture is to convince you, a general
physics audience, of two things:

{\ }\quad a) There already {\em is} today a fairly complete theory of quantum
geometry, at least as a new mathematical language.  There are several
approaches
and the one I describe here and which I consider fairly complete is my own one
which I called the {\em braided approach} (or {\em braided
geometry}\cite{Ma:bra}\cite{Ma:introp}).  There are braided lines\cite{Ma:any},
planes\cite{Ma:poi}, matrices\cite{Ma:exa}\cite{Ma:add}, differentials and
plane
waves (exponentials)\cite{Ma:fre}, Gaussians and integration\cite{KemMa:alg},
forms and epsilon tensors\cite{Ma:eps} and more known in this approach in
association which a general $R$-matrix, i.e.  a general solution of the
celebrated quantum Yang-Baxter equations (QYBE).  These equations are usually
associated with the quantum groups of Faddeev, Jimbo and
Drinfeld\cite{FRT:lie}\cite{Dri}\cite{Jim:dif} and these do indeed play a role
in the braided approach as a background covariance of our constructions, but
not
as the geometry itself.  I will begin with the more conventional (but not so
complete) approach in which quantum groups are geometrical objects too, and
then
move on the systematic braided theory.  This approach is quite different also
from the approach to non-commutative geometry comming out of the work of A.
Connes and others.

{\ }\quad b) This new language of quantum or braided geometry does indeed make
it possible to formulate new and previously unsayable concepts in theoretical
physics.  The duality principle which is the title of my lecture will be such a
new concept.  This in turn allows the possibility to realise this new concept
experimentally and thereby to discover new and previously inconceivable
physical
phenomena.  This is what I mean by the deep interdependence of theoretical
physics and pure mathematics. One can develop this point of view into a Kantian
or Hegelian perspective on the concept of reality in physics\cite{Ma:pri}.

The new and previously unsayable physical phenomenon which I want to
concentrate
on here is a generalisation of wave-particle duality or position-momentum
symmetry.  I would like to formulate this mathematically as a consequence of
something a bit deeper, which I call a {\em quantum-geometry transformation}.
This is the assertion that the symmetry algebra (or generalised momentum) of a
system could also be isomorphic to the coordinate algebra of some geometry.
That geometry would be momentum space as a geometrical object.  This is
certainly possible in flat space where \eqn{U=C}{ U(\R^n)=\C[\vecp]=\C(\R^n).}
Here $U(\R^n)$ is the enveloping algebra of the momentum generators
$\vecp=\{p^i\}$ which is the symmetry point of view, while $\C(\R^n)$ is the
algebra of coordinate functions $\{p^i\}$ on momentum space as a geometrical
object.  They are obviously isomorphic since both are polynomials in
$n$-variables.  In fact, this isomorphism is as {\em Hopf algebras}.  The
axioms
of a Hopf algebra will be given next in Section~2 but they have a coproduct map
$\Delta$ which on the left is trivial (on the left the Lie algebra structure of
$\R^n$ is reflected in the product of the enveloping algebra).  On the right
the
same generators are viewed as coordinates and have a coproduct which
corresponds
to addition in momentum space, while the algebra is the trivial one.  So the
isomorphism is perhaps remarkable.  The right hand side is the way that we will
always deal with geometry in this lecture, in terms of the algebra of
coordinates.  It is a geometrical picture of momentum against a picture as
differential operators ${\del\over\del x_i}$.

The algebra $\C(\R^n)=\C[\vecx]$ as generated by position coordinates $\{x_i\}$
is how we think of position space.  So (\ref{U=C}) allows us to think of the
momentum in the same language as we think of position, but with $\vecp$ in the
role of $\vecx$.  We can also say it in the reverse way, \eqn{C=U}{
C(\R^n)=\C[\vecx]=U(\R^n)} whereby we think of $x_i={\del\over\del p^i}$ as
differential operators on momentum space.  This is wave-particle duality.

As well as being isomorphic, the position and momentum are connected in a
symmetrical way by this differentiation.  This can be formulated as the
assertion that the two Hopf algebras $\C[\vecx]$, $\C[\vecp]$ are dually paired
by \eqn{evalxp}{ \<\phi(\vecp),\psi(\vecx)\>=(\phi(\del)\psi)(0)} making them
dual as Hopf algebras.  An element of one defines a linear functional on the
other.  We will see the precise definition in Section~2 but it should be clear
that the dual Hopf algebra to $\R^n$ is $\hat{\R}^n\isom\R^n$ again because
$\R^n$ is a self-dual Abelian group.  A basis and dual basis are \[
\phi^{m_1\cdots m_n}=(p^1)^{m_1}\cdots (p^n)^{m_n},\quad \psi_{m_1\cdots
m_n}={x_1^{m_1}\cdots x_n^{m_n}\over m_1!  m_2!\cdots m_n!}.\] The map
(\ref{evalxp}) is the evaluation $\C[\vecp]\tens \C[\vecx]\to \C$ as dual
linear
spaces.  But there is also a canonical element called the {\em coevaluation}
and
defined simply as the element in $\C[\vecx]\tens\C[\vecp]$ corresponding to the
identity map $\C[\vecx]\to\C[\vecx]$ in the usual way since $\C[\vecp]$ is dual
to $\C[\vecx]$.  In our case it is a formal powerseries (rather than living in
the algebraic tensor product) and comes out as \eqn{coevxp}{\coev=\sum_{m_i}
\psi_{m_1\cdots m_n}\tens \phi^{m_1\cdots m_n}=e^{\vecp\cdot\vecx}.}  This is a
rather modern way of thinking about exponentials and Fourier theory, but one
that generalises well to quantum geometry\cite{KemMa:alg}.

So conceptually, $\C[\vecp]$ and $\C[\vecx]$ are dual to each other, but they
are also isomorphic (i.e.  self-dual).  This is our formulation of
wave-particle
duality.  It works just as well for any self-dual Abelian group (such as
$\Z_n$).  It also works when the group is not self-dual but still Abelian.
Then
the Fourier conjugate space is the dual group, e.g.  $\hat{\Z}=S^1$.  The dual
point of view with position and momentum interchanged is no longer isomorphic
but both are still possible and in a dual relationship.

On the other hand, this duality phenomenon completely fails to get off the
ground when the symmetry or generalised momentum group is non-Abelian.  If
$\cg$
is a non-Abelian Lie algebra and we ask for a geometrical picture \eqn{UgC}{
U(\cg)=\C(?)} as functions on some ordinary manifold, we can never succeed
simply because the left hand side is non-commutative, while the pointwise
multiplication of functions on a manifold on the right hand side is always
commutative.  So for a geometrical picture we need some more general
non-commutative or quantum geometry.  To give an elementary example, consider
the sphere $S^2$ with a standard Poisson bracket.  Its Kirillov-Kostant
quantisation is \cmath{[x_i,x_i]=\lambda \eps_{ij}{}^k x_k,\quad \sum_i
x_i^2=1.}  This is just $U(su_2)$ modulo a `constant distance' relation.  So at
least in this case we can think geometrically of $U(su_2)$ as a quantum space
even though the $x_i$ are non-commuting coordinates.  Moreover, with the
greater
generality of non-commutative or quantum geometry we can continue to pursue
our position-momentum symmetry or duality phenomenon.

\subsection*{Acknowledgements}

This work was prepared at the Erwin Schr\"odinger Institute in Vienna.  I would
like to thank the staff there for their encouragement and support.  It was
presented here in Istanbul and I thank the organisers at the Bosphorus
University for their excellent hospitality as well.

\section{What IS a Hopf algebra or quantum group?}

Probably the reader has seen the definition of a Hopf algebra or quantum group
elsewhere.  There are several points of view which, remarkably, all lead to the
same set of axioms.  Here I want to give the axioms from the duality point of
view, which is the one that interests us here.

In this case a Hopf algebra $A$ is just an algebra equipped with further
structure such that the dual linear space $A^*$ of functionals on $A$ is also
an
algebra in a compatible way.  In terms of $A$ this additional structure is a
{\em coproduct} $\Delta:A\to A\tens A$.  It looks like a product but goes in
the
opposite direction.  While an algebra is of course associative, the coproduct
is
{\em coassociative}.  These conditions are 
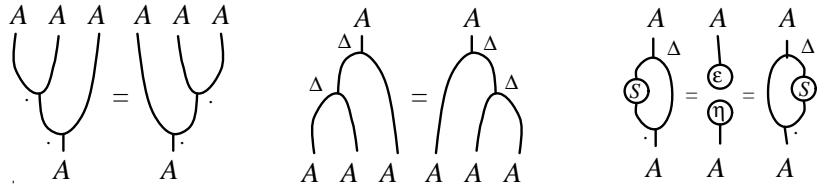
\[ \epsfbox{ass-coass.eps}\] where
we
also show axioms for an {\em antipode} or `linearised inverse' map $S:A\to A$.
In addition, we require that $\Delta$ is an algebra homomorphism.  We also
require a {\em counit} map $\eps:A\to \C$ obeying
$(\eps\tens\id)\circ\Delta=\id=(\id\tens\eps)\circ\Delta$.

The nice thing about these axioms is their input-output symmetry:  Turn them
up-side-down and you get the same axioms with the roles of $\cdot$ and $\Delta$
interchanged!  The counit axioms become the ones for the inclusion of the
identity element $\C\to A$ which we usually take for granted.  This
input-output
symmetry is a deep feature of quantum geometry in the form that comes out of
Hopf algebras.  For our present purposes we realise it as follows:  {\em If $A$
is a Hopf algebra} (say finite-dimensional) {\em then so is $A^*$}.  The
structures correspond according to \[ \<\phi\psi,a\>=\<\phi\tens\psi,\Delta
a\>,\quad \<\phi,ab\>=\<\Delta\phi,a\tens b\>,\quad \<\phi,Sa\>=\<S\phi,a\>\]
for all $\phi,\psi\in A^*$ and $a,b\in A$.  In the non-finite dimensional case
we can still look for such a {\em duality pairing} between two Hopf algebras.

Thus $\C[\vecx]$ and $\C[\vecp]$ are Hopf algebras which are dually paired in
this way.  The Hopf algebra structure is \[ \Delta x_i=x_i\tens 1+1\tens
x_i,\quad \eps x_i=0,\quad Sx_i=-x_i\] and similarly for $\vecp$.  For a
general
function $a(\vecx)$ the coproduct $\Delta a(\vecx)$ is a function of two
variables with value $(\Delta a)(x,y)=a(x+y)$.  So the coproduct expresses the
addition law on $\R^n$ in terms of the co-ordinate functions.  The addition law
is of course the basis for the geometry of $\R^n$.

More generally if $G$ is a suitably nice group there will be a Hopf algebra
$\C(G)$ generated by coordinates on $G$ and again with a coproduct
corresponding
to the group law.  If the group is non-Abelian the geometry here typically has
curvature.  It corresponds to the output of $\Delta$ not being symmetric or
cocommutative.  We also have another Hopf algebra $U(\cg)$ generated by the Lie
algebra of $G$.  This is dual \[ U(\cg)=\C(G)^*\] in the loose sense that we
have a duality pairing between these Hopf algebras.  Since enveloping algebras
are typically part of the algebra of quantum observables of a quantum system,
we see that Hopf algebras suggest a general principle that {\em quantum theory
and geometry are in a dual relationship}, implemented in the simplest case by
Hopf algebra duality.

When we look away from commutative examples such as $\C(G)$ we can still think
of a Hopf algebra as a generalisation of the functions on a group manifold,
even
when there is no manifold in the usual sense.  Likewise, when we look away from
cocommutative examples such $U(\cg)$ we can still think of a Hopf algebra as
like an enveloping algebra or a Lie algebra.  Because the axioms of a Hopf
algebra are self dual, one algebraic concept contains both points of view.
This
means that the category of Hopf algebras unifies two concepts in physics:  one
linked to quantum mechanics and the other to geometry.  This is why the
language
of Hopf algebras opens up the possibility to solve (\ref{UgC}) by generalising
both sides slightly and why we call it a {\em quantum-geometry transformation}.

This also suggests that Hopf algebras are a natural category in which to
develop
some simple models in which quantum theory and gravity are unified.  This is
the
approach to Planck scale physics introduced in the author's
thesis\cite{Ma:the}\cite{Ma:phy}\cite{Ma:pla}\cite{Ma:pri} on the basis of the
above duality considerations.  Obviously a natural source of non-commutative
algebras $A$ is as quantum algebras of observables, which are non-commutative
versions of the classical algebra of functions on phase space.  So we ask:

\begin{itemize} \item When is a quantum algebra of observables a Hopf algebra?
\item If so, what does it mean physically?  \end{itemize}

The answer to the first question is that it happens quite often; demanding it
is
a non-trivial but interesting constraint to put on a physical system.  I will
describe a large class of such physical systems associated to certain
homogeneous spaces\cite{Ma:the}\cite{Ma:pla}.  The answer to the second
question
is two fold:

{\ }\quad (a) We now have geometry of phase space (in the crude form of a group
law) even in the quantum setting, i.e.  a unification of quantum theory and
geometry.  A general quantum system would include in its algebra of observables
the enveloping algebra $U(\cg)$ for any generalised momentum generators; so to
render such an algebra as like functions on a manifold involves for the
momentum
part exactly the quantum-geometry transformation (\ref{UgC}).  We now have
non-commutativity coming from the momentum-position cross relations which we
also have to deal with quantum geometry.

{\ }\quad (b) We have the possibility of a new kind of symmetry in physics in
which we reinterpret $A^*$ as the algebra of observables of a dual system, and
$A\subset A^{**}$ as containing the states of the dual system.  Thus we think
of
\[ \<\phi,a\>=\phi(a)=\sum_i \rho_i\<\phi_i|a|\phi_i\>\] as the expectation
value of observable $a$ in mixed state $\phi=\sum\rho_i\<\phi_i|\ |\phi_i\>$
while the dual system considers the same number as $\<\phi,a\>=a(\phi)$, the
expectation value of the dual-observable $\phi$ in the dual-state $a$.  Note
that we work with states in the usual way in mathematics as (say positive)
linear functionals on the algebra of observables.  These are typically convex
linear combinations of expectations against Hilbert-space states, as indicated
here.  Thus we have the possibility of a second physical system with
observables
$A^*$ dual to our original one.  This is a rather radical phenomenon in the
quantum world which we propose to call {\em observable-state} or {\em
micro-macro} duality\cite{Ma:pla}\cite{Ma:phy}.  It is also connected as we
have
seen with input-output symmetry, a point which is developed further in
\cite{Ma:wal}.

The simplest example in 1 dimension is the Hopf algebra $\C[x]\bicross \C[p]$
generated by $x,p$ with the relations and Hopf
structure\cite{Ma:phy}\cite{Ma:clau} \cmath{[p ,x]=-{A\over B}(1-e^{-Bx}),\quad
\Delta x=x\tens 1+1\tens x,\quad \Delta p= p\tens e^{-Bx}+1\tens p\\ \eps
x=\eps
p=0,\quad Sx=-x,\quad Sp=-pe^{Bx}} depending on two real parameters $A,B$.  It
is the most general solution of the problem in 1-dimensions within the
bicrossproduct construction in the next section.  If we identify
$\hbar=-{A\over
B}$ we have a deformation of usual quantum mechanics.  If we keep $p$ as a
conserved momentum and keep Hamiltonian ${-p^2\over 2m}$ so that the particle
is
in free-fall then the modified commutation relation results in modified
dynamics.  We use conventions in which $p$ is antihermitian, i.e.  ${-\imath
p\over m}$ is the velocity.  Then \[ {dx\over dt}={\imath\over\hbar}[-{p^2\over
2m},x]=({-\imath p\over m})(1-e^{-Bx})+ O(\hbar),\quad {dp\over dt}=0.\] Hence
if we consider a particle falling in from $\infty$ we see that as the particle
approaches the origin $x=0$ it goes more and more slowly.  In fact, it takes an
infinite amount of time to reach the origin, which therefore behaves in some
ways like a black-hole event horizon.  This analogy should not be pushed too
far
since our present treatment is in nonrelativistic quantum mechanics, but it
gives us an estimate $B={c^2\over M\grav}$ as introducing the distortion in the
geometry comparable to a gravitational mass $M$.  Here $\grav$ is Newton's
constant.

On the other hand we get a commutative algebra if we take the limit $\hbar\to
0$.  In this case $\C[x]\bicross \C[p]\isom \C(\R\rcross\R)$ where $\R\rcross
\R$ has a non-Abelian group law corresponding to the coproduct above.  It has
curvature related to $B$.  Combining with the above we see that our Hopf
algebra
has two limits\cite{Ma:phy}\cite{Ma:clau}:

\begin{picture}(300,50)(-10,-15) \put(0,0){$\C[x]\bicross
\C[p]$}\put(55,8){\vector(3,1){28}}\put(55,0){\vector(3,-1){28}}
\put(86,13){$\C[x]\lcross \C[p]$ usual quantum mechanics} \put(86,-13){$\C(X)$
usual curved geometry} \put(14,19){\sevrm $mM\ll m_P^2$}\put(14,-19){\sevrm
$mM\gg m_P^2$} \end{picture}

\bigskip \noindent where $m_P$ is the Planck mass.  In the first limit the
particle motion is not detectably different from usual quantum mechanics
outside
the Compton wavelength from the origin.  Of course, there is still a
singularity
in our model right at the origin.  In the second limit the non-commutativity
would not show up for length scales larger than the background gravitational
scale set by $M$.  This demonstrates our goal of unifying quantum mechanical
and
gravitational effects within a single model.

The dual Hopf algebra is generated by linear functionals $\phi,\psi$ defined by
\[ \<\phi, :a(x,p):\>=({\del a\over \del x} )(0,0),\quad
\<\psi,:a(x,p):\>=({\del a\over \del p} )(0,0)\] where $:a(x,p):=\sum a_{n,m}
x^np^m$ is the normal-ordered form of a function in the two variables $x,p$.
Then we have \cmath{\nquad [\psi,\phi]=\hbar^{-1}(1-e^{-A\psi}),\quad \Delta
\phi=\phi\tens 1+e^{-A\psi}\tens\phi, \quad \Delta \psi=\psi\tens
1+1\tens\psi\nonumber\\ \eps\phi=\eps\psi=0,\quad S\phi=-e^{A\psi}\phi,\quad
S\psi=-\psi} which is of just the same type as the above:  \[ (\C[x]\bicross
\C[p])^*=\C[\phi]\cobicross \C[\psi]\isom \C[p]\cobicross \C[x].  \] So this
particular Hopf algebra is self-dual.  It means that the observable-state
duality is indeed realised in this model, the dual quantum system being of a
similar form.  This is a new kind of symmetry principle which one can propose
as
a speculative idea for the structure of Planck scale physics\cite{Ma:pla}.

\section{Bicrossproduct Hopf algebras}

The example at the end of the last section is one of a large class of
bicrossproduct models\cite{Ma:pla} which we describe now.  We consider first
the
finite group case and then move on to the Lie version.

The data for the bicrossproduct construction is a {\em group factorisation}.
This means a group $X$ with subgroups $G,M\subset X$ such that $G\times M\to X$
given by multiplying within $X$ is an isomorphism.  In this case multiplying in
the group and projecting down to $M$ and $G$ gives mutual actions \[ \ra:
M\times G\to M ,\quad \la:M\times G\to G\] from the right and left
respectively,
obeying \ceqn{mpair}{ s\ra e=s,\ (s\ra u)\ra v=s\ra(uv);\quad e\ra u=e,\
(st)\ra
u=\left(s\ra(t\la u)\right)(t\ra u)\\ e\la u=u,\ s\la(t\la u)=(st)\la u;\quad
s\la e=e,\ s\la(uv)=(s\la u)\left((s\ra u)\la v\right)} for all $u,v\in G$ and
$s,t\in M$.  The first two in each line say that we have an action, and the
second say that the action is almost by automorphisms, but twisted by the other
action.

This data is just what it takes to obtain a Hopf algebra $\C(M)\bicross\C G$
built on the vector space with basis $G\times M$.  We write the basis elements
as labelled squares $s{\ \atop{\lform\atop \displaystyle u}}=s{\displaystyle
s\la u \atop { \lform\atop \displaystyle u}}s\ra u$ where the convention is
that
if we label the left and lower edges then the other two are labelled by the
values transformed by the actions $\la,\ra$.  In this notation, the Hopf
algebra
structure is\cite{Ma:clau} \[ \epsfbox{bichopf.eps}\] The product consists of
gluing the squares horizontally whenever the edges are suitably matched.  The
coproduct by contrast consist of {\em ungluing} vertically, i.e.  it is the sum
of all pairs of squares which when glued vertically would give the square we
began with.  The dual Hopf algebra is $(\C(M)\bicross\C G)^*=\C
M\cobicross\C(G)$ and has just the same form with vertical gluing and
horizontal
ungluing.  The roles of $\la,\ra$ and $G, M$ are interchanged.

In mathematical terms the algebra here is a {\em cross product algebra}.  This
is a more or less standard way to quantise particles moving under the action of
a group\cite{Mac:ind}\cite{DobTol:mec}.  In nice cases the action $\ra$ of $G$
on $M$ induces a metric with the particle then moving on geodesics.  The Lie
algebra $\cg$ of $G$ plays the role of momentum since its elements generate the
geodesics or 1-parameter flows.  This works for any action $\ra$.  The main
result of \cite{Ma:the} is that if in this setting of particles on homogeneous
spaces there is a {\em backreaction} $\la$ obeying (\ref{mpair}) then the
resulting quantum algebra of observables is a Hopf algebra.  Not every action
$\ra$, i.e.  not every homogeneous space, admits such a backreaction, i.e.  it
is a strong constraint on the system.  We have seen the flavour of the
constraint in the example $\C[x]\bicross\C[p]$:  it forces non-linear dynamics.
One can think of it within this class of models as some kind of `Einstein's
equation'\cite{Ma:pla} since it is an (integrated) second order constraint on
$\ra$.  We see then that the duality principle which leads us to look for a
Hopf
algebra structure of this type forces something a bit like Einstein's equation.

Note that we gave the Hopf algebras above for finite groups, while for physics
we need to work with Lie groups and topological spaces.  This can be done with
Hopf-von Neumann algebras\cite{Ma:hop} as well as algebraically in the form
$\C(M)\bicross U(\cg)$ with dual $U(\cm)\cobicross \C(G)$ where $\cg,\cm$ are
the respective Lie algebras.  There are examples for all $\cg$ compact
semisimple, with $\cm=\cg^{\star\rm op}$ a suitable solvable Lie algebra coming
from the Iwasawa decomposition\cite{Ma:mat}.

For the simplest 3-dimensional example we take $\cg=su_2$ and
$M=SU(2)^{\star\rm
op}$.  Then $\C(SU(2)^{\star\rm op})\bicross U(su_2)$ is generated by position
co-ordinates $\{\xpos_1,\xpos_2,\xpos_3,(\xpos_3+1)^{-1}\}$ and $su_2$
generators $\{e_1,e_2,e_3\}$ with \cmath{ \nquad [\xpos_i,\xpos_j]=0, \quad
\Delta \xpos_i=\xpos_i\tens 1+(\xpos_3+1)\tens \xpos_i, \ \eps \xpos_i=0,\quad
S\xpos_i=-{\xpos_i\over \xpos_3+1}\\ {} [e_i,e_j]=\eps_{ij}{}^ke_k,\quad
[e_i,\xpos_j]=\eps_{ij}{}^k\xpos_k-{\eps_{ij3}\over 2}{\xpos^2\over
\xpos_3+1},\quad \eps e_i=0\\ \Delta e_i=e_i\tens {1\over \xpos_3+1}+e_3\tens
{\xpos_i\over \xpos_3+1}+1\tens e_i,\quad Se_i=e_3\xpos_i-e_i(\xpos_3+1).}

That this is a Hopf algebra is rather hard to check directly, and is best
done via the theory above.  We start with the action \[ e_i\la
\xpos_j=\eps_{ij}{}^k(\xpos_k-{\delta_{k3}\over 2}{\xpos^2\over \xpos_3+1})\]
of
$su_2$ on $\C(SU(2)^{\star\rm op})$ induced by $\ra$ above, and then make a
cross product $\C(SU(2)^{\star\rm op})\lcross U(su_2)$ to obtain the algebra.
For the coalgebra we follow the dual construction:  we construct from $\la$ in
(\ref{mpair}) a {\em coaction} \[ \beta(e_i)=e_i\tens {1\over
\xpos_3+1}+e_3\tens {\xpos_i\over \xpos_3+1}\] and make a {\em cross coproduct}
construction $\C(SU(2)^{\star\rm op})\rcocross U(su_2)$ to obtain $\Delta$.
The
axioms for a coaction $\beta$ here are like the axioms for an action but with
the arrows reversed.  For experts, there is also a $*$-structure $e_i^*=-e_i$
and $\xpos_i^*=\xpos_i$ making the above into a Hopf $*$-algebra.  We gave the
Hopf-von Neumann algebra version in \cite{Ma:hop}.

The manifold $SU(2)^{\star\rm op}$ is the region $\{\vecs\in\R^3,\quad s_3>
-1\}$ equipped with a deformed (non-Abelian) group
law\cite{Ma:pla}\cite{Ma:mat}.  The action $\ra$ is a deformation of the usual
action of $su_2$ by rotations.  Its orbits are non-concentrically nested
spheres
in this region, accumulating as $s_3=-1$.  Particles quantised on these orbits
are described by the above Hopf algebra.  It is a deformed quantum top.  The
dual system consists of the Lie algebra $su_2^{\star\rm op}$ acting on the
group
manifold $SU(2)$ with orbits which are in fact the symplectic leaves in $SU(2)$
for the Sklyanin bracket, i.e.  an equally interesting system.

Recently, a four dimensional example similar to this one has been found in
\cite{MaRue:bic} with $M$ a non-Abelian version of $\R^{1,3}$ and
$\cg=so(1,3)$.
Writing $\{\xpos_\mu\}$ for the position-coordinates and $\{ N_i,M_i\}$ for the
Lorentz generators we have the Hopf algebra $U(so(1,3))\cobicross \C(M)$ as
\cite{MaRue:bic} \cmath{ \nquad [\xpos_\mu,\xpos_\nu]=0,\quad
[M_i,M_j]=\eps_{ij}{}^kM_k,\quad [ N_i, N_j]=\eps_{ij}{}^k N_k,\quad [M_i,
N_j]=0\\ {} [\xpos_0,M_i]=0,\quad [\xpos_i,M_j]=\eps_{ij}{}^k \xpos_k\\
{}\nquad\ [\xpos_0, N_i]=-\xpos_i,\quad [\xpos_i,
N_j]=-\delta_{ij}\left({\kappa\over
2}(1-e^{-{2\xpos_0\over\kappa}})+{\xpos^2\over 2\kappa}\right)+{\xpos_i
\xpos_j\over\kappa}\\ \Delta \xpos_0=\xpos_0\tens 1+1\tens \xpos_0,\quad \Delta
\xpos_i=\xpos_i\tens 1+e^{-{\xpos_0\over\kappa}}\tens \xpos_i\\ \nquad \Delta
M_i=M_i\tens 1+1\tens M_i,\quad \Delta N_i= N_i\tens
1+e^{-{\xpos_0\over\kappa}}\tens N_i+{\eps_i{}^{jk}\over\kappa} \xpos_j\tens
M_k} along with a suitable counit, antipode and $*$-structure.  According to
the
above, we consider this the quantum algebra of observables of a system
consisting of particles moving in orbits in our curved $\R^{1,3}$ under a
deformed Lorentz transformation.

On the other hand, it is obvious that both this example and the preceding
$\R^3$
example from \cite{Ma:pla}\cite{Ma:mat}\cite{Ma:hop} could be considered
instead
as deformed enveloping algebras of 3-dimensional or 4-dimensional Poincar\'e
Lie
algebras \[ \C(SU(2)^{\star\rm op})\bicross U(su_2)\equiv U_\kappa({\rm
p}_3),\quad U(so(1,3))\cobicross \C(M)\isom U_\kappa({\rm p}(1,3)).\] In the
first case we introduce a parameter $\kappa$ in the formulae above.  In the
second case we recover a Hopf algebra isomorphic\cite{MaRue:bic} to the
$\kappa$-deformed Poincar\'e Hopf algebra introduced by other means in
\cite{LNRT:def}.  We regard the $\xpos_i$ or $\xpos_\mu$ generators now as
momentum generators.  This demonstrates once again the quantum geometry
transformation (\ref{UgC}), this time dualising the interpretation of the
position co-ordinates to view them as momentum.

\section{Bicrossproduct Poisson structures and Lie bialgebras}

This section announces new material in which we look at the above ideas at the
semiclassical level.  The semiclassical notion of a Hopf algebra is a {\em
Poisson-Lie group}.  At the Lie algebra level it is a {\em Lie
bialgebra}\cite{Dri:ham}.  These notions are useful in the theory of classical
inverse scattering where Lie algebra splittings lead to solutions of the
Classical Yang-Baxter equations.  The examples we give in this section are
however, quite different from this standard theory:  they are by contrast the
actual Poisson brackets whose quantisation is the bicrossproduct quantum
algebras of observables described in the last section.  The latter are likewise
far removed from the usual quantum groups $U_q(\cg)$ of Drinfeld and Jimbo,
being a different origin of quantum groups and Hopf algebras as coming out of
ideas for Planck scale physics in \cite{Ma:pla}.  Further details of the
results
in this section will be given in \cite[Chapter~8]{Ma:book}.

Briefly, a Lie bialgebra is a Lie algebra $\cg$ such that $\cg^*$ is also a Lie
algebra in a compatible way.  In terms of $\cg$ it means that there is an
additional structure $\delta:\cg\to\cg\tens\cg$ called the {\em Lie cobracket}
and required to obey \eqn{liebialg}{\nquad\delta=-\tau\circ\delta,\quad
(\id\tens\delta)\circ\delta\xi +{\rm cyclic}=0,\quad
\delta([\xi,\eta])=\ad_\xi(\eta)-\ad_\eta(\xi)} for all $\xi,\eta\in \cg$.
Here
$\tau$ is the usual transposition and $\ad$ is the adjoint action extended to
$g\tens g$ as a derivation.  The dual of a finite-dimensional Lie bialgebra is
also a Lie bialgebra with the role of $[\ ,\ ]$ and $\delta$ interchanged.

The last condition in (\ref{liebialg}) can be understood as Lie algebra
cocycle.
Hence if $G$ is a connected simply connected Lie group with Lie algebra $\cg$
then $\delta$ exponentiates to a a group cocycle $D:G\to g\tens g$.  This in
turn defines a Poisson bracket on $G$ by \[ \{f,g\}(u)=(R_{*}D)(f\tens
g)(u),\quad \forall f,g\in C^\infty(G)\] where $R_{*u}$ is the right
translation
$\cg=T_eG\to T_uG$ applied to the tensor square $g\tens g$ to give the
tensor-field $R_{*}D$.  We evaluate this 2-tensor field as a differential
operator on the functions $f$ and $g$.  This is the geometrical meaning of a
Lie
bialgebra\cite{Dri:ham}.  It corresponds to a class of Poisson brackets on the
group manifold.  The Poisson brackets in this class behave well with respect to
the group product.

We find such a structure now in the semiclassical part of our bicrossproduct
Hopf algebras.  These are quantisations of $C^\infty(X)$ where $X$ is the
classical phase space and in our case is a Lie group corresponding to the
coproduct structure.  The commutator in our quantum algebra of observables to
lowest order in $\hbar$ gives us the Poisson bracket.  To describe the result
let us start with the bicrossproduct data (\ref{mpair}) in Lie algebra form.
This data is a pair $\cg,\cm$ of Lie algebras acting on each other by \[
\ra:\cm\tens\cg\to\cm ,\quad \la:\cm\tens\cg\to\cg \] obeying the matching
conditions\cite{Ma:phy} \ceqn{liempair}{[\phi,\psi]\la\xi=\phi\la(\psi\la\xi) -
\psi\la(\phi\la\xi),\quad
\phi\ra[\xi,\eta]=(\phi\ra\xi)\ra\eta-(\phi\ra\eta)\ra\xi\\ {}
\phi\la[\xi,\eta]
=[\phi\la\xi,\eta]+[\xi, \phi\la\eta]+(\phi\ra\xi)\la\eta-(\phi\ra\eta)\la\xi\\
{} [\phi,\psi]\ra\xi=[\phi\ra\xi,\psi]+[\phi,
\psi\ra\xi]+\phi\ra(\psi\la\xi)-\psi\ra(\phi\la\xi)} for all $\xi,\eta\in\cg$
and $\phi,\psi\in\cm$.

It was shown in \cite{Ma:mat} that subject to a completeness condition for some
resulting vector fields, such Lie algebras acting on each other exponentiate to
Lie groups as in (\ref{mpair}).  Moreover, we have such a pair whenever a Lie
algebra splits as a vector space into sub-Lie algebras $\cg,\cm$.  This bigger
Lie algebra is reconstructed as a {\em double cross sum} $g\dcross m$ it is
built on $\cg\oplus \cm$ with the Lie bracket in \cite{Ma:phy}. It is generated
by $\cg,\cm$ as sub-Lie algebras and the cross-bracket \[ [\phi,\xi]=\phi\la\xi
+ \phi\ra\xi .\]
The construction can be useful for generating more and more solutions of
(\ref{liempair}) in view of the proposition:

\begin{propos} Let $(\cg,\cm,\ra,\la)$ be a Lie algebra matched pair as in
(\ref{liempair}).  Then $(\cg^{\rm op}\oplus\cm ,\cg\dcross\cm)$ is also a Lie
algebra matched pair, where $\cg\dcross\cm$ acts by the left adjoint action and
$\cg^{\rm op}\oplus\cm $ acts by $-\ra$ for the action of $\cg^{\rm op}$ and
$-\la$ for the action of $\cm$.  \end{propos} Explicitly, $\cg^{\rm op}$ is
$\cg$ with the negated Lie bracket.  The action of $\cg\dcross \cm$ and
matching
`backreaction' of $\cg^{\rm op}\oplus\cm$ are
\cmath{(\xi\oplus\phi)\la(\eta\oplus\psi)=([\xi,\eta]+\phi\la\eta-\psi\la\xi)\oplus
([\phi,\psi]+\phi\ra\eta-\psi\ra\xi)\\
(\xi\oplus\phi)\ra(\eta\oplus\psi)=(-\psi\la\xi)\oplus(-\phi\ra\eta)} when both
are built on the vector space $\cg\oplus\cm$,  using the Lie brackets of
$\cg,\cm$ and the original actions.  These new actions obey (\ref{liempair})
for
our bigger system.  We then obtain a new Lie algebra $(\cg^{\rm
op}\oplus\cm)\dcross(\cg\dcross\cm)$, a construction which can clearly be
iterated.

By contrast to these double cross sum Lie algebras, we now obtain from the same
data (\ref{liempair}) a Lie bialgebra.

\begin{propos} Let $(\cg,\cm,\ra,\la)$ be a Lie algebras matched pair as in
(\ref{liempair}).  There is a {\em bicross-sum} Lie bialgebra $\cm^*\bicross
\cg$ generated by $\cm^*$ with zero Lie bracket, $\cg$ and the cross relations
and Lie coalgebra \cmath{ [\xi, f]=\xi\la f,\quad \delta f=\<f,[e_a,e_b]\>
e^a\tens e^b,\quad \delta \xi=e_a\la\xi\tens e^a-e^a\tens e_a\la\xi} for all
$f\in \cm^*$ and $\xi\in\cg$.  Here $\{e_a\}$ is a basis of $\cm$ and
$\{e^a\}$ a dual basis.  \end{propos}

This is built as a vector space on $\cm^*\oplus\cg$ with the Lie bracket etc.,
defined by the above on the two components.  The proof is a matter of some
detailed calculations to check the axioms (\ref{liebialg}) for a Lie bialgebra.
To explain this construction in detail one has to introduce the notion of Lie
bialgebra coactions $\beta$ dual to the notion of a Lie algebra action.  Then
the Lie bracket is the usual semidirect sum Lie bracket for the action of $\cg$
on $\cm^*$ induced by $\ra$, while the cobracket $\delta$ is a {\rm
co-semidirect sum} by a Lie coaction $\beta$ induced by $\la$.  The dual Lie
bialgebra is \[ (\cm^*\bicross \cg)^*=\cm\cobicross \cg^*\] built in an
analogous way as another bicross-sum with the roles of $\cm,\cg$ interchanged.

These constructions, like $\cg\dcross\cm$, all have generalisations to the case
when $\cg,\cm$ are Lie bialgebras to begin with.  This more general theory is
the semiclassical part of the double cross product and bicrossproduct of
general
Hopf algebras in \cite[Sec.  3]{Ma:phy}.  We do not try to cover it here, but
see \cite{Ma:book}.  Instead, we give some examples of Proposition~4.2 and its
dual.

For the first bicrossproduct example in Section~3, the Lie algebra data is as
follows.  We have $\cg=su_2=\{e_i\}$ and $\cm=su_2^{\star\rm op}=\{x^i\}$ say,
with \eqn{su2*}{[x^1,x^2]=0,\quad [x^3,x^1]=x^1,\quad [x^3,x^2]=x^2 }
\eqn{su2*su2}{ x^i\ra e_j=\eps^i{}_{jk} x^k,\quad x^i\la
e_j=e_3\delta^i{}_j-\delta^i{}_3e_j} which solve (\ref{liempair}).  Both Lie
algebras here are 3-dimensional and $\cm\cobicross\cg^*$ is six-dimensional.
It
has the structure (\ref{su2*}) and \cmath{[e^i,e^j]=0,\quad
[x^i,e^j]=\delta^i{}_3 e^j-e^i\delta^j{}_3\\ \delta e^i=\eps^i{}_{jk}e^j\tens
e^k,\quad \delta x^i=\eps^i{}_{jk}(e^j\tens x^k-x^k\tens e^j)} where $\{e^i\}$
is a dual basis to the $su_2$ basis.  Its associated Lie group is a
six-dimensional manifold $X$ on which the cobracket $\delta$ defines a natural
Poisson structure as explained above.  Its quantisation is the bicrossproduct
Hopf algebra in Section~3.  From another point of view, the same Hopf algebra
is
a deformation of the enveloping algebra of a 3-dimensional Poincar\'e Lie
bialgebra $\cm^*\bicross\cg$ with structure from Proposition~4.2 given by
\cmath{ [\xpos_i,\xpos_j]=0,\quad [e_i,e_j]=\eps_{ij}{}^k e_k,\quad
[e_i,\xpos_j]=\eps_{ij}{}^k\xpos_k\\ \delta \xpos_i=\xpos_3\tens
\xpos_i-\xpos_i\tens \xpos_3,\quad \delta e_i=e_3\tens\xpos_i-\xpos_i\tens
e_3+\xpos_3\tens e_i-e_i\tens\xpos_3} where $\{\xpos_i\}$ is a dual basis to
the
$\{x^i\}$.

The same theory applies to the example at the end of Section~3.  We have
$\cg=so(1,3)=\{M_i,N_j\}$ and $\cm=\{x^\mu\}$ say, with\cite{MaRue:bic}
\eqn{kappamink}{[x^i,x^j]=0,\quad [x^i,x^0]=x^i} \ceqn{minkbic}{M_i\ra
x^0=M_i\ra x^j=0,\quad N_i\ra x^0=-N_i,\quad N_i\ra x^j=\eps_i{}^{jk}M_k\\
\nquad\nquad M_i\la x^0=0,\quad M_i\la x^j=\eps_i{}^j{}_k x^k,\quad N_i\la
x^0=-\delta_{ij}x^j,\quad N_i\la x^j=-\delta_i{}^j x^0.}  One can check that
$(\cm,\cg,\ra,\la)$ is again a Lie algebra matched pair.  The Lie bialgebra
$\cg^*\bicross\cm$ is ten-dimensional and has the structure from
Proposition~4.2
given by (\ref{kappamink}) and
\cmath{{}[M^i,M^j]=[N^i,N^j]=[M^i,N^j]=[x^i,N^j]=[x^0,M^i]=0\\
{}[x^0,N^i]=-N^i,\quad [x^i,M^j]=\eps^{ij}{}_k N^k,\quad \delta
N^i=\eps^i{}_{jk}N^j\tens N^k\\ \delta M^i=\eps^i{}_{jk}M^j\tens M^k,\quad
\delta x^0=(N^i\tens x^j-x^j\tens N^i)\delta_{ij}\\ \delta x^i=N^i\tens
x^0-x^0\tens N^i+\eps^i{}_{jk}(M^j\tens x^k-x^k\tens M^j)} where $\{N^i,M^i\}$
are a dual basis to the Lorentz basis.  Its associated Lie group is a
ten-dimensional manifold $X$ on which the cobracket $\delta$ defines a natural
Poisson structure as explained.  Its quantisation is the bicrossproduct Hopf
algebra at the end of Section~3.  From another point of view, the same Hopf
algebra is a deformation of the enveloping algebra of a 4-dimensional
Poincar\'e
Lie bialgebra $\cg\cobicross\cm^*$ with structure \cmath{ \nquad\nquad
[\xpos_\mu,\xpos_\nu]=[\xpos_0,M_i]=[M_i, N_j]=0,\quad
[M_i,M_j]=\eps_{ij}{}^kM_k,\ [ N_i, N_j]=\eps_{ij}{}^k N_k\\ {}
[\xpos_i,M_j]=\eps_{ij}{}^k\xpos_k,\quad
[\xpos_i,N_j]=-\delta_{ij}\xpos_0,\quad
[\xpos_0,N_i]=-\xpos_i\\ \delta\xpos_0=0,\quad
\delta\xpos_i=\xpos_i\tens\xpos_0-\xpos_0\tens\xpos_i\\ \nquad\nquad\delta
M_i=0,\quad\delta N_i=N_i\tens\xpos_0-\xpos_0\tens
N_i+\eps_i{}^{jk}(\xpos_j\tens M_k-M_k\tens\xpos_j)} where $\{\xpos_\mu\}$ is a
dual basis to the $\{x^\mu\}$.

\section{Braided Geometry and $q$-Minkowski space}

Now we outline another and more complete approach to quantum geometry which has
emerged recently in \cite{Ma:bra}\cite{Ma:exa} and subsequent works by the
author and collaborators.  In this approach the deformation of usual geometry
is
not connected conceptually with quantisation, but with {\em braid statistics}.
We want to explore our Hopf algebra duality ideas in this context as well.

The idea of this {\em braided geometry} is to start with super-geometry and
replace the its usual Bose-Fermi statistics $\pm 1$ by something more general.
This can be an arbitrary factor $q$ \cite{Ma:csta} or more generally it can be
a
matrix solution $R$ of the quantum Yang-Baxter equation (QYBE).  This equation
is connected historically with quantum groups\cite{FRT:lie} and quantum inverse
scattering, but we will not use it in this historical way.  I.e., this is a new
approach to $q$-deformation.  There are already two big reviews by the
author\cite{Ma:introp}\cite{Ma:varen} so we shall be brief.

The starting point for braided geometry is the concept of a {\em braided
group}\cite{Ma:bra}.  In mathematical terms a braided group $B$ is defined like
a Hopf algebra or quantum group as in Section~2, but the coproduct
$\und\Delta:B\to B\tens B$ is no longer multiplicative as it was there.
Instead, it is {\em braided-multiplicative} in the sense \eqn{brahom}{
\und\Delta(ab)=(\und\Delta a)(\und\Delta b),\quad (a\tens c)(b\tens
d)=a\Psi(c\tens b)d} for all $a,b,c,d\in B$.  In diagrammatic form we use an
operator $\Psi=\epsfbox{braid.eps}:B\tens B\to B\tens B$ called the braiding in
those places where diagrammatically we would need to write a braid crossing,
i.e.  in those places where there is a transposition in our constructions.  We
are working in fact in a kind of braided mathematics.  The familiar case on
which we model everything is the super case where \[ \Psi(c\tens b)=b\tens c
(-1)^{|c||b|}\] on homogeneous elements of degree $|\ |$.

The simplest new example is the braided line $B=\C[x]$ with structure \[
\nquad\und\Delta x=x\tens 1+1\tens x,\quad\und\eps x=0,\quad \und S x=-x,\quad
\Psi(x^m\tens x^n)=q^{mn}x^n\tens x^m\] The new part is $\Psi$ and means for
example that \[ \nquad \und\Delta x^m=\sum_{r=0}^m [{m\atop r};q] x^r\tens
x^{m-r},\quad [{m\atop r};q]={[m;q]!\over [r;q]![m-r;q]!},\quad
[m;q]={1-q^m\over 1-q}\] which is the origin in braided geometry of the
$q$-integers $[m,q]$ familiar in working with $q$-deformations.

The next simplest example is the braided plane $B$ generated by $x,y$ with
\cmath{ yx=qxy,\quad \und\Delta x=x\tens 1+1\tens x,\quad \und\Delta y=y\tens
1+1\tens y\\ \und\eps x=\und\eps y=0,\quad \und Sx=-x,\quad \und Sy=-y\\
\Psi(x\tens x)=q^2 x\tens x,\quad \Psi(x\tens y)=q y\tens x,\quad \Psi(y\tens
y)=q^2 y\tens y \\ \Psi(y\tens x)=q x\tens y+(q^2-1)y\tens x} The algebra here
is sometimes called the `quantum plane'; the new part is the coproduct
$\und\Delta$ and the braiding $\Psi$.  The latter is the same one that leads to
the Jones knot polynomial in a more standard context.

The general construction of which these are examples is associated to a pair of
invertible matrices $R,R'\in M_n\tens M_n$ obeying
\cmath{R_{12}R_{13}R_{23}=R_{23}R_{13}R_{12}\\
R_{12}R_{13}R'_{23}=R'_{23}R_{13}R_{12},\quad
R'_{12}R_{13}R_{23}=R_{23}R_{13}R'_{12}\\ (PR+1)(PR'-1)=0,\quad R_{21}R'
=R'_{21}R} where $P$ is the permutation matrix and $R_{12}=R\tens \id$ in
$M_n^{\tens 3}$, etc.  Given such data we define the associated braided
covector
algebra $B$ with generators $\{x_i\}$ as\cite{Ma:poi} \cmath{
x_ix_j=x_bx_aR'{}^a{}_i{}^b{}_j,\quad {\rm i.e.,}\quad
\vecx_1\vecx_2=\vecx_2\vecx_1R' \\ \und\Delta x_i=x_i\tens 1+1\tens x_i,\quad
\und\eps x_i=0\,\quad \und S x_i=-x_i\\ \Psi(x_i\tens x_j)=x_b\tens x_a
R^a{}_i{}^b{}_j,\quad {\rm i.e.,}\quad \Psi(\vecx_1\tens \vecx_2)=\vecx_2\tens
\vecx_1R .}  We use here and below a shorthand notation where the suffices on
bold-face vectors etc., refer to the position of the indices in a tensor
product
of matrices.

We can proceed to develop geometry in this general setting, starting with this
$\und\Delta$, which we call `coaddition' since it has the additive form.  We
can
bring this out by using a notation $\vecx\equiv\vecx\tens 1$,
$\vecx'\equiv1\tens\vecx$.  In our braided tensor product in (\ref{brahom})
these two do not commute but rather obey the braid-statistics \[
\vecx'_1\vecx_2=\vecx_2\vecx'_1R\] corresponding to $\Psi$.  The
braided-homomorphism property of $\und\Delta$ is then just the statement that
$\vecx''=\vecx+\vecx'$ obey the same relations of $B$ provided we use these
braid statistics between our two independent copies.

In this notation, we define braided-differentiation as\cite{Ma:fre} \[ \del^i
f(\vecx)= \left(a_i^{-1}(f(\veca+\vecx)-f(\vecx))\right)_{\veca=0}\equiv{\rm
coeff\ of\ }a_i{\rm \ in\ } f(\veca+\vecx)\] and find that these operators obey
the relations $\del_1\del_2=R'\del_2\del_1$ of a {\em braided vector algebra}.
This is defined like our covectors $B$ but with upper indices.  We have a a
braided-Leibniz rule \[ \del^i(bc)=(\del^i b)c+\cdot\Psi^{-1}(\del^i\tens
b)c,\quad\forall b,c\in B\] where there is a natural braiding between vectors
and covectors defined also by $R$.  In braided geometry all independent objects
enjoy braid statistics with respect to each other.

On monomials the braided differentiation comes out as \[ \del^i(\vecx_1\cdots
\vecx_m)= {\bf e}^i{}_1\vecx_2\cdots\vecx_m \left[m;R\right]_{1\cdots m}\]
where
${\bf e}^i$ is a basis covector $({\bf e}^i){}_j=\delta^i{}_j$ and
\[\left[m;R\right]=1+(PR)_{12}+(PR)_{12}(PR)_{23} +\cdots+(PR)_{12}\cdots
(PR)_{m-1,m}\] is a certain {\em braided integer matrix}\cite{Ma:fre}.  In the
1-dimensional case the latter become the usual $[m,q]$ and $\del^i$ becomes the
celebrated Jackson $q$-derivative.  Using the braided integers we can define
braided-binomial coefficient matrices\cite{Ma:fre} from which the coproduct of
$x_{i_1}\cdots x_{i_m}$ can be recovered much as in the 1-dimensional case
above.

It is also possible to define a braided-exponential or `plane wave' following
the same lines as in (\ref{coevxp}):  we define a pairing between the braided
covectors and vectors of the form (\ref{evalxp}) with $\del$ the braided one,
and define $\exp$ to be the corresponding coevaluation as a formal powerseries.
It is an eigenfunction of the $\del^i$ as well as an eigenfunction with respect
to braided differentiation $\del\over\del p^i$.  We have a braided
wave-particle
duality.  Note that if the braided integer matrices are all invertible then
$\exp$ takes a simple form using $([m;R]!)^{-1}$ in the powerseries.  In the
one-dimensional case we recover the celebrated $q$-exponential
$\sum_{m=0}^\infty{x^m p^m\over [m,q]!}$.

We can also define a braided Gaussian\cite{KemMa:alg} as the solution $g_\eta$
of the equation \[ \del^i g_\eta=-x_a \eta^{ai}g_\eta,\quad
\und\eps(g_\eta)=1\]
again as a formal powerseries.  Here $\eta^{ij}$ is a {\em braided metric}
defined in such a way that $x_a\eta^{ai}$ behaves like a braided vector.  It
obeys a number of identities with $R,R'$. If $\eta$ obeys some further
identities, the Gaussian takes a nice form involving the 1-dimensional
$q$-exponential of $x_jx_i\eta^{ij}$.

One can also define translation-invariant integration $\int$.  More precisely,
it turns out to be more natural to define a linear functional $\CZ:B\to \C$
where\cite{KemMa:alg} \[\CZ[f(\vecx)]=\left(\int
f(\vecx)g_\eta\right)\left(\int
g_\eta\right)^{-1};\quad \CZ[x_i]=0,\quad
\CZ[x_ix_j]=\eta_{ba}R^a{}_i{}^b{}_j\lambda^2\] etc.  Here $\lambda$ is a
certain constant depending on $R$, the quantum group normalisation
constant\cite{Ma:lin}.

Finally, we add to our assumptions on $R$ the additional equations \cmath{
R_{12}R'_{13}R'_{23}=R'_{23}R'_{13}R_{12},\quad
R'_{12}R'_{13}R_{23}=R_{23}R'_{13}R'_{12} \\
R'_{12}R'_{13}R'_{23}=R'_{23}R'_{13}R'_{12}} which gives a kind of symmetry
$R\swap -R'$.  Using it, we can define forms $\theta_i=\extd x_i$ as like the
above braided covectors but with $-R$ in place of $R'$.  This leads to the
relations \cmath{ \extd\vecx_1\extd\vecx_2=-\extd\vecx_2\extd\vecx_1R,\quad
\vecx_1\extd\vecx_2=\extd\vecx_2\vecx_1 R\\ \extd (\extd x_{i_1}\cdots\extd
x_{i_p} f(\vecx))=\extd x_{i_1}\cdots\extd x_{i_p} \extd x_a {\del\over\del
x_a}
f(\vecx)} for exterior differentials in the braided approach.  As this point we
make contact with similar formulae for $\extd x_i$ imposed in other
approaches\cite{WesZum:cov} by consistency arguments.  We also define the
epsilon tensor as\cite{Ma:eps} \[ \eps^{i_1 i_2\cdots i_n}={\del\over\del
\theta_{i_1}}\cdots {\del\over\del
\theta_{i_n}}\theta_1\cdots\theta_n=([n;-R']!)^{i_n \cdots i_1}_{12\cdots n}\]
where we braided-differentiate in form-space.

This is our survey of braided geometry.  We also have a natural candidate in
this approach for braided-Minkowski space.  We just use the braided matrices
$B(R)$ introduced in \cite{Ma:exa} with generators $\{u^i{}_j\}$ and relations
$R_{21}\vecu_1R\vecu_2=\vecu_2R_{21}\vecu_1R$.  These relations can be put in
the form of a braided covector space with $u_I=u^{i_0}{}_{i_1}$ in a
multi-index
notation and $\bf R'$ a suitable matrix built from $R$.  We did this in
\cite{Ma:exa} while the correct $\bf R$ for the additive braiding $\Psi$ was
found by U.  Meyer in \cite{Mey:new} and corresponds to the braid statistics
$R^{-1}\vecu_1'R\vecu_2= \vecu_2 R_{21}\vecu_1'R$.  It assumes that $R$ obeys a
certain Hecke condition.  There is also a third matrix $\bf R_\cdot$
corresponding to a different braiding $\Psi$ needed for a braided group with
multiplicative $\und\Delta \vecu=\vecu\tens\vecu$ as introduced in
\cite{Ma:exa}.  In short, we have natural matrix-like braided groups with both
coaddition and comultiplication.

Moreover, for $R$ obeying a certain reality condition, there is a natural
$*$-structure defined by $u^i{}_j{}^*=u^j{}_i$, i.e. our braided matrices are
hermitian\cite{Ma:mec}.  Then it is obvious that $2\times 2$ braided hermitian
matrices using, for example, the standard Jones polynomial R-matrix are a good
definition of $q$-Minkowski space.  It has four generators
$\vecu=\pmatrix{a&b\cr c&d}$ and relations \cmath{ba=q^2ab,\quad
ca=q^{-2}ac,\quad d a=ad,\qquad bc=cb+(1-q^{-2})a(d-a)\\ d
b=bd+(1-q^{-2})ab,\quad cd=d c+(1-q^{-2})ca.}  Previously
\cite{CWSSW:lor}\cite{OSWZ:def} proposed a similar algebra as $q$-Minkowski
space on the basis of tensoring two copies of the quantum plane as spinors.
The
braided approach on the other hand means that we get all the structure above:
braid statistics, coaddition, differentiation, exponentiation, Gaussians,
integration, forms and the epsilon tensor.  We refer to the literature and the
review \cite{Ma:varen} for details.  See also\cite{MaMey:bra}\cite{Mey:wav}.

Finally, we return to the braided theory and ask about braided-Lie algebras and
braided-enveloping algebras.  There is such a theory introduced by the author
and based on the axioms\cite{Ma:lie} \[\epsfbox{Liefrag.eps}\] where our
braided
Lie-algebra $\CL$ has a bracket $[\ ,\ ]$ and an additional `sharing out' map
$\Delta$ which we usually take for granted when working with ordinary Lie
algebras.  There is also a map $\eps$.  We refer to \cite{Ma:lie} for details
and for the theorem that every braided-Lie algebra has a universal enveloping
braided group $U(\CL)$.  Also in \cite{Ma:lie} is a general class of examples
of
the form \cmath{ \CL=\span\{u^i{}_j\},\quad \Delta \vecu=\vecu\tens\vecu, \quad
[\vecu_1,R\vecu_2]= R_{21}^{-1}\vecu_2 R_{21}R\\ \Psi(R^{-1}\vecu_1 \tens
R\vecu_2)=\vecu_2 R^{-1}\tens \vecu_1 R.}  One has $U(\CL)=B(R)$ for this
braided-Lie algebra.  We can also generate it by the generators
\[\chi^i{}_j=u^i{}_j-\delta^i{}_j,\quad
R_{21}\chi_1R\chi_2-\chi_2R_{21}\chi_1R=R_{21}R\chi_2-\chi_2R_{21}R.\] Again
the
Jones polynomial R-matrix can be fed into this construction and gives us the
braided-Lie algebra $gl_{2,q}$.  It has basis $h,x_+,x_-,\gamma$ with
braided-Lie bracket \align{&&[h,x_+]=(q^{-2}+1)q^{-2}x_+=-q^{-2}[x_+,h]\\
&&[h,x_-]=-(q^{-2}+1)x_-=-q^{2}[x_-,h],\quad [x_+,x_-]= q^{-2}h=-[x_-,x_+]\\
&&[h,h]=(q^{-4}-1)h, \quad [\gamma,\cases{h\cr x_+\cr
x_-}]=(1-q^{-4})\cases{h\cr x_+\cr x_-}} and zero for the others.  We see that
as $q\to 1$ the $\gamma$ mode decouples and we have the Lie algebra $su_2\oplus
u(1)$, but for $q\ne 1$ these are unified.  On the other hand, the enveloping
bialgebra $U(\CL)$ comes out as the isomorphism\cite{Ma:lie} \[
U(gl_{2,q})\isom
\R_q^{1,3},\quad \pmatrix{h\cr x_+\cr x_-\cr\gamma}=(q^2-1)^{-1}\pmatrix{a-d\cr
c \cr b\cr q^{-2}a+d-(q^{-2}+1)}\] which is another example of a
quantum-geometry transformation, now in our braided setting!  What is
remarkable
about this is that when $q\to 1$ the left hand side is the enveloping algebra
of
$su(2)\oplus u(1)$ and is non-commutative, but the right hand side is the
coordinate algebra of Minkowski space and becomes commutative, i.e.  the
isomorphism is only possible in the braided $q\ne 1$ world!  Such an
isomorphism
between two of the key ingredients of the standard model in particle physics
suggests a deep application of q-deformation in this context.

\end{document}